\documentclass[]{aa}  
\usepackage{graphicx}
\usepackage{txfonts}
\usepackage{natbib}
\usepackage{dcolumn}
\usepackage{subfigure}
\usepackage{hyperref}
\usepackage{epstopdf}
\usepackage{dcolumn}
\usepackage{url}

\newcommand{\RNum}[1]{\uppercase\expandafter{\romannumeral #1\relax}}
\bibpunct{(}{)}{;}{a}{}{,} % to follow the A&A style
\begin{document}
   \title{Period change of massive binaries from combined photometric and spectroscopic data in Cygnus OB2}
	\titlerunning{period change of binaries from combined data in Cygnus OB2}

   \author{Jaan Laur\inst{1,2}
          \and
	  Elmo Tempel\inst{1}
                \and	
          Taavi Tuvikene\inst{1}
                \and	
	  T\~onis Eenm\"ae\inst{1}
		\and	
	  Indrek Kolka\inst{1}
	  }

   \institute{
             Tartu Observatory, Observatooriumi~1, 61602 T\~oravere, Estonia\\
             \email{jaan.laur@to.ee}
          \and
             Institute of Physics, University of Tartu, Ravila 14c, 51010 Tartu, Estonia}
	  % \fnmsep
   \date{Received May 12, 2015; accepted June 29, 2015}

  \abstract
   {Mass loss is an important property in evolution models of massive stars. As up to 90\% of the massive stars have a visual or spectroscopic companion and many of them exhibit mass exchange, mass-loss rates can be acquired through the period study of massive binaries.}%context
   {Using our own photometric observations as well as archival data, we look for variations in orbital periods of seven massive eclipsing binary systems in the Cygnus OB2 association and estimate their mass-loss rates and stellar parameters.}%aims
   {We use a Bayesian parameter estimation method to simultaneously fit the period and period change to all available data and a stellar modelling tool to model the binary parameters from photometric and radial-velocity data.}%methods
   {Four out of the seven selected binaries show non-zero period change values at two-sigma confidence level. We also report for the first time the eclipsing nature of a star MT059.}%results
   {}%conclusions

   \keywords{binaries: eclipsing -- stars: mass-loss -- open clusters: individual: Cygnus OB2 -- stars: massive}

   \maketitle
%
%________________________________________________________________

\section{Introduction}

The change of a period ($\dot{P}$) in a binary system has been commonly evaluated by measuring the times of minimum of primary eclipses and then comparing those times to the calculated ephemerides for the eclipses.
The period change value is then acquired by fitting a polynomial model through the observed minus calculated ($O-C$) data points.
The value of this period change can indicate mass loss through stellar winds or mass transfer between the binary components \citep{Singh:1986}.

Mass loss in binary systems comes mostly from the mass transfer via Roche lobe overflow when the primary star's photosphere crosses the inner Lagrange point and starts to exchange mass with its companion.
While the mass is transferred, the centre of mass of the interacting binary system shifts towards the mass-receiving star. 
Due to the conservation of angular momentum, this mass transfer causes a change in the system's orbital period: the period increases while the mass is transferred to the more massive star and vice versa.
The period also increases when mass is lost from the system through stellar winds.

\citet{Sana:2012} studied a sample of massive O stars and found that over 70\% of them exchange mass with a companion.
The speckle interferometric survey of Galactic massive stars \citep{Mason:2009} expanded upon a decade-old survey and noted that 75\% of the massive stars in clusters and 59\% in the field have either a spectroscopic or a visual companion.
As low inclination or long periods make the binary systems hard to detect, these values are only the lower limit and, when corrected to the observational bias, the upper limit for binarity fraction in bound systems can reach as high as $\sim~90~\pm~10\%$ \citep{Kiminki:2012_2}.

Massive stars have a relatively short lifetime that is influenced by their mass-loss rates.
This mass loss determines the evolution of their temperature and luminosity, and also the supernova explosion in the final evolution phases.
Massive stars provide feedback in the form of ultraviolet (UV) radiation, stellar winds, and supernovae which impact the evolution of disc galaxies by mixing and driving turbulence of interstellar gas and triggering star formation.
They also enrich the interstellar medium with heavy elements and drive the chemical evolution of galaxies and the metallicity evolution of the universe.
Therefore, knowing the evolution of massive stars gives us insight into the evolution of galaxies.
The uncertainties in determining the mass-loss rates for massive stars are still large even in the local Universe \citep{Smith:2014}.

Constraining orbital parameters for massive binary systems is important to test the predictions of theoretical models.
In order to determine all the parameters of a single binary system, one needs to obtain photometric and spectroscopic data at various orbital phases, requiring great observational effort.
As it is time-consuming to obtain those observations and as the eclipsing nature is dependent on the inclination of the system, there are not many massive binaries with fully constrained parameters known.
In this work we present the stellar parameters as well as the period and the period change values of seven selected systems in Cygnus OB2.

Stellar association Cygnus OB2 is one of the most massive and richest associations known in our Galaxy, containing at least 60 O-type stars \citep{Negueruela:2008}.
Determined from the stellar evolution models, the age of the cluster is found to be 1--7~Myr \citep{Wright:2015}.
It is dynamically young, sub-structured and unbound association that is located in the molecular cloud complex Cygnus X and its total estimated mass is $(1.6 \pm 0.3)~\times~10^4~M_{\odot}$ \citep{Wright:2014}.

The Cygnus OB2 Radial Velocity Survey (RVS) of 128 selected O and early-B stars is underway \citep{Kiminki:2007,Kiminki:2008,Kiminki:2009,Kiminki:2012,Kiminki:2014} to assess the binary properties of massive stars within a common formation environment at a similar age. 
There are now 48 massive binary systems known in Cygnus OB2.

We have chosen Cygnus OB2 to study the mass loss in massive binaries because of the previously published data for massive stars in that association.
Based on the radial-velocity (RV) data and our own photometric data that we obtained during 2011--2014, we have chosen seven massive binary systems with eclipsing light curves.
In order to assess the change of period in those systems, we use a Bayesian parameter estimation method.
We also use binary modelling software to acquire orbital and physical parameters.
For several stars that were previously modelled using RV data only, we present models with additional photometric data.

%----------------------------------------------------------
\section{Data}

In this work we combine new photometric observations with photometry and radial-velocity data from the literature.

\begin{table}
\caption{References to all the data that is used in our period change analysis.}
\centering
\label{datafrom}
\begin{tabular}{lccc}
\hline \hline
\rule{0pt}{10pt}Star & RA& DEC &References \\
 & Hour & Deg & \\
 \hline 
\rule{0pt}{10pt}\object{A36} &20:34:59 &+41:36:17& 1,2,3 \\
\object{B17} &20:30:27& +41:13:25& 1,2\\
\object{MT059} & 20:31:11 & +41:31:54& 1,2,4,5\\
\object{MT696} &20:34:00&+41:17:35& 1,4,6,7\\
\object{MT720} &20:34:06 &+41:08:09& 1,4,5\\
\object{Schulte 3} &20:31:38&+41:13:21& 1,2,4,5,8\\
\object{Schulte 5} & 20:32:22& +41:18:19& 1,2,4,8,9,10,\\
&&& 11,12,13,14,15,16\\

\hline
\end{tabular}
\tablebib{
(1)~iTelescope (our data set 1);
(2) NSVS\footnote{\url{http://skydot.lanl.gov/nsvs/nsvs.php}};
(3) \citet{Kiminki:2009};
(4) RAITS (our data set 2);
(5) \citet{Kiminki:2008};
(6) \citet{Souza:2014};
(7) \citet{Kiminki:2012};
(8) IOMC\footnote{\url{https://sdc.cab.inta-csic.es/omc/index.jsp}};
(9) \citet{Sazonov:1961};
(10) \citet{Hall:1974};
(11) HIPPARCOS\footnote{\url{http://www.rssd.esa.int/SA/HIPPARCOS/apps/PlotCurve.html}};
(12) \citet{Linder:2008};
(13) \citet{Miczaika:1953};
(14) \citet{Bohannan:1976};
(15) \citet{Rauw:1999};
(16) \citet{Wilson:1951}.
}

\end{table}

References to the data that we used for each star are given in Table~\ref{datafrom}.
We use the same naming conventions as in the Cygnus OB2 RVS papers: the MT names from \citet{Massey:1991}, Schulte from \citet{Schulte:1958} and A36 and B17 from \citet{Comeron:2002}.
The data sets are described in more detail below.

We monitored Cygnus OB2 with two telescopes during 2011--2014.
The first data set was obtained with the commercial iTelescope.net 0.25-m Takahashi Epsilon telescope (T4), located in Mayhill, New Mexico, USA.
An SBIG ST-10 CCD camera with Johnson-Cousins $BV\!I$ filters was used, providing a field of view (FOV) of $60.5~\times~40.8$ arcmin.
Observations were carried out on 344 nights during 2011--2013.
An observation consisted of three consecutive exposures in each filter.
Data reduction was carried out using the PHOTWORK software \citep{Tuvikene:2012}.
Magnitudes were extracted by means of aperture photometry, except for MT720 that required PSF photometry due to a neighbouring star.
To minimise observational errors, we used the nightly mean instrumental differential photometry values \citep[for details see][]{Laur:2012}.

The second data set was obtained on 103 nights during two seasons, 2013--2014, using the Tartu Observatory 0.31-m RAITS telescope, located in T\~oravere, Estonia.
The telescope was equipped with an Apogee Alta U42 CCD camera and Johnson-Cousins $BV\!RI$ filters.
The FOV was $37~\times~37$ arcmin.
Five consecutive exposures were taken with each filter during an observation.
The CCD frames were reduced and magnitudes were extracted by aperture photometry as described above.

\begin{figure}[t]
	\includegraphics[width=0.49\textwidth, trim=0 0 0 0,clip]{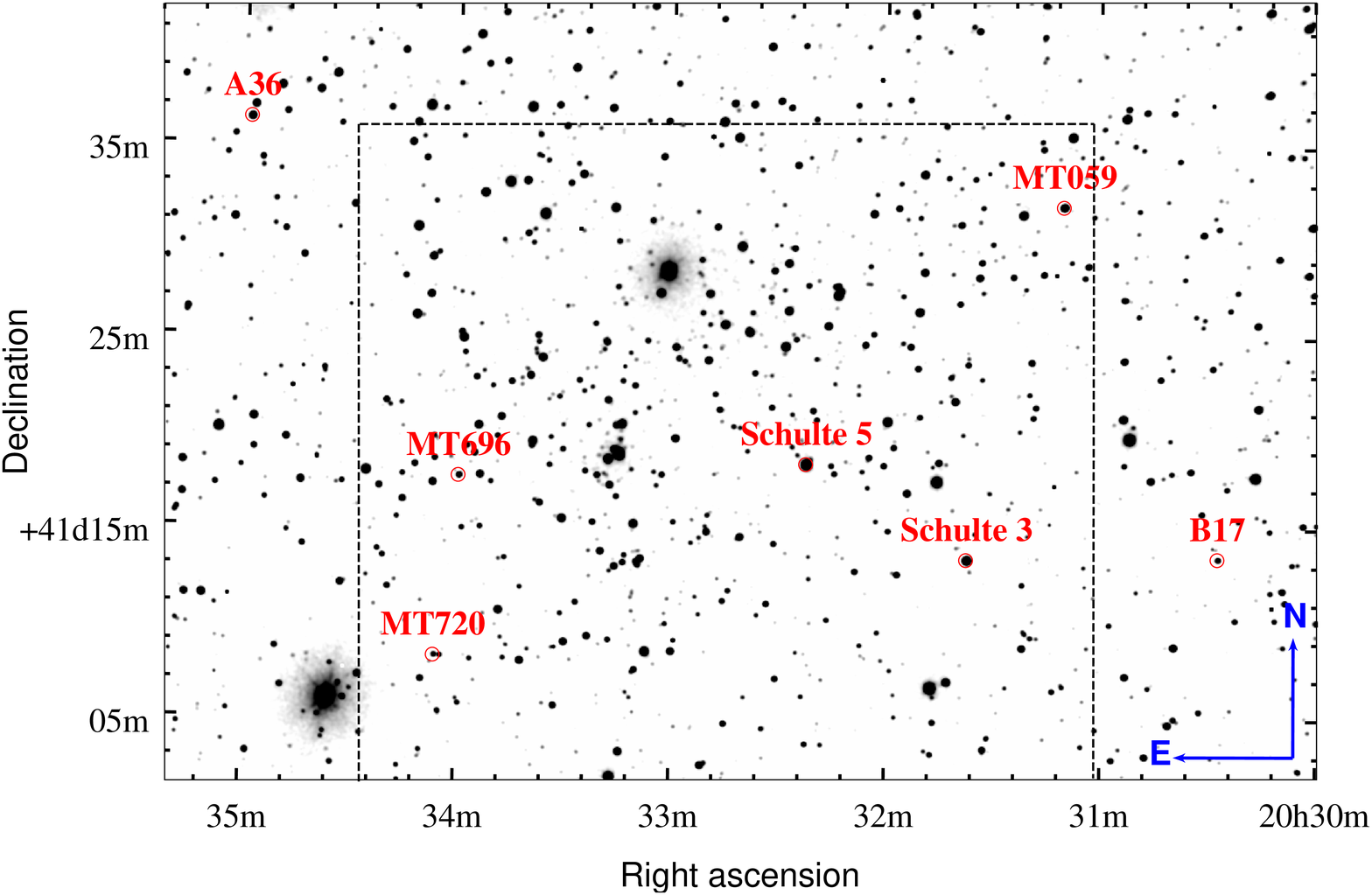}
	\caption{The full frame of Cygnus OB2 region from our first data set (iTelescope) with the selected binaries shown in red. The dashed line indicates the FOV of our second data set (RAITS).}
	\label{CygOB2-ds9}
\end{figure}

Figure~\ref{CygOB2-ds9} shows the FOV of the two telescopes we used.
Two of the seven program stars, A36 and B17, fell outside of the RAITS FOV.

The Cygnus OB2 RVS is a spectroscopic survey of unbiasedly chosen 128 stars of which 83 are B-type and 45 are O-type stars.
They have used Keck+HIRES spectrograph (1999--2001), the Lick+Hamilton echelle spectrograph (1999--2000), the WIYN+Hydra spectrograph (2001--2008) and a longslit optical spectrograph with a 2.3-m telescope at the Wyoming Infrared Observatory (2011--2013, 2014 May). 
The Cygnus OB2 RVS data that we used, ranges from 1999 to 2009.
The longest data set of 43 measurements (1999--2007) is for MT059.  
Schulte~3 and A36 have the least observations ($\sim$15) and a time base of one year while MT720 have 32 observations in two years and MT696 have 31 measurements with the time base of 30 days.

The Northern Sky Variability Survey~(NSVS) covers the entire northern sky in an unfiltered optical region, ranging from 8 to 15.5 magnitudes from the Robotic Optical Transient Search Experiment (ROTSE-I) \citep{Wozniak:2004}.
Except for MT696, there are $170$ data points obtained within one year (April -- December, 1999) for all of our selected stars.
The data set for MT720 was available but due to large photometric errors, the signal of its eclipses was unattainable and we had to exclude it from our analysis.
The errors for NSVS data sets are typically larger than our own data sets because their data quality is limited by a severe blending due to 14-arcsec pixel size.

The Hipparcos catalogue data \citep{VanLeuween:2007} are available only for the brightest star in our selection, Schulte~5, because of the limiting magnitude of Hipparcos observations.
The available data span over three years, from the end of 1989 to the beginning of 1993, with hundred broad-band measurements.

The Optical Monitoring Camera (OMC) on board of the ESA International Gamma-Ray Astrophysics Laboratory (INTEGRAL) mission consists of $5^{\circ}\times$~$5^{\circ}$ FOV photometric camera observing in the Johnson $V$-band \citep{MasHesse:2003}.
The only stars in our selection that have been observed by IOMC are Schulte~3 and Schulte~5.
There have been more than 5000 observations made between 2002 and 2009 (which is our longest time series from a single source) and their optical accuracy is roughly on the same level with NSVS photometry.

\citet{Souza:2014} have been monitoring MT696 since 2010 in two 5-nm wide filters centred on the continuum and $H\alpha$ (645 and 656 nm respectively).
Their photometric data span for over three years from 2010 to the end of 2013 and consist of 105 observations.
These data and \citet{Kiminki:2012} RV data are the only published time series of MT696 to date.

For Schulte~5 we have the longest time base with measurements from 12 different sources.
The furthest published measurements for Schulte~5 are the times of minimum from 1898 although the accuracy of those measurements are under suspicion by the authors themselves \citep{Sazonov:1961}.
During a 5-month period in 1952, \citet{Miczaika:1953} obtained 36 measurements with a photoelectric equipment on the 32-cm refractor at the Heidelberg Observatory.
\citet{Hall:1974} observed Schulte~5 photoelectrically for 142 times with a 60-cm Seyfert reflector during the 1967 to 1970 period using a set of closely matching $U\!BV$ passband filters.
\citet{Linder:2009} acquired over 300 exposures during a 14-night campaign in 1998 with five filters of which three were centred on different emission-line wavelengths found in various subsets of Wolf-Rayet objects and other two were centred on line-free areas (around $5057$ and $6051$~\r{A}).
For our period change analysis we chose only the measurements taken with the continuum narrowband filters as the emission line data also indicate variability not directly related to the eclipses of the binary.

Radial-velocity data for Schulte~5 are available from three sources \citep{Wilson:1951,Bohannan:1976,Rauw:1999}.
All the observations span three years (1946--1949, 1971--1974 and 1994--1997) and consist of 42, 7 and 25 measurements accordingly.
Schulte~5 has the longest time base for RV data as it is the only binary in our selection that has had more than one RV observation campaign carried out.

\begin{figure}[t]
	\includegraphics{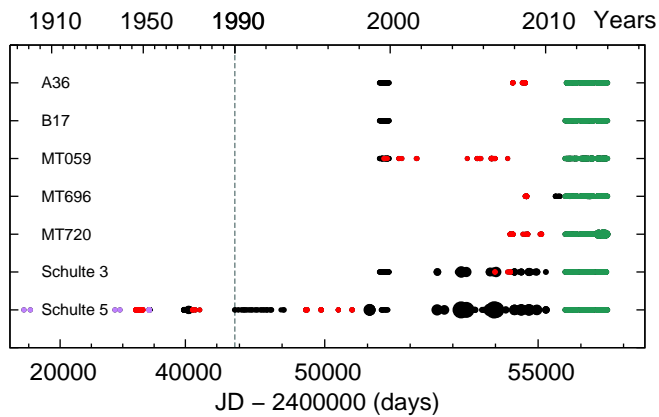}
	\caption{The available data shown in time domain with reduced Julian dates at the bottom and calendar years at the top. Point sizes are indicating the number of measurements in a one-day bin. We have changed the scale of the plot in the middle (dashed line) to better show the data available from 1990 onwards. Green points indicate our photometric data sets, red points are the available radial-velocity data, purple points show the times of minimum from \citet{Sazonov:1961} and black points are all other available photometric data listed in Table \ref{datafrom}.}
	\label{aegrida}
\end{figure}

All the available data points are shown in Fig.~\ref{aegrida} to illustrate the time base for each star.
It can be seen from the figure that Schulte~5 is by far the most observed star in our selection whereas the other massive binaries have been scarcely studied.
Our analysis depends mostly on three sources: our own two data sets from 2011 to 2014, NSVS and Cygnus OB2 RVS.
While Schulte~5 has a time base of 116 years, all the other binaries have a shorter time base than 15 years. 

\begin{figure}[t]
	\includegraphics{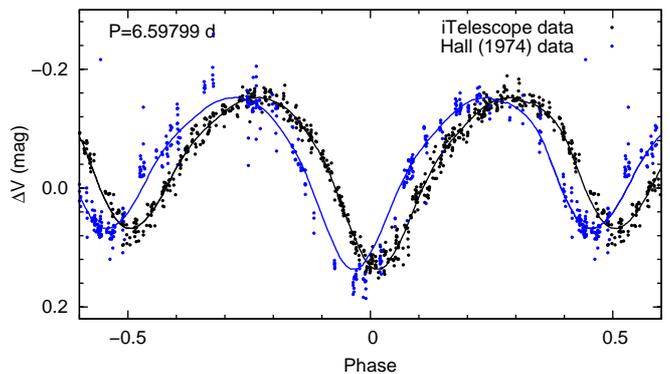}
	\caption{The $V$-filter phase diagram of Schulte~5, folded with a constant period. The iTelescope data (black points) and \citet{Hall:1974} data (blue points) have been obtained 45 years apart. Solid lines are smoothed empirical light curves. The visible phase shift is due to the cumulative effect of period change ($\dot{P}$).}
	\label{hall_itel}
\end{figure}

The time base is important in period change studies as a constant change in period introduces a gradual shift in eclipse timings for each consecutive orbit.
This causes a measurable cumulative phase shift over a long time base.
The longer the time base, the more apparent the shift becomes, increasing the accuracy of the period change assessments.
Figure~\ref{hall_itel} illustrates the cumulative effect of the period change for Schulte~5 with two different data sets that are observed 45 years apart.

\section{Method}\label{sec:Method}

To estimate the variations in orbital period of our selected massive binary systems, we used a Bayesian approach of simultaneously fitting all available observational data: light curves, times of minimum and RV data.

This method consists of two distinctive steps: acquiring the model curves and then assessing the period ($P$) and period change ($\dot{P}$) values that simultaneously fit the data to those curves (Table \ref{results_period}).
We assume that the shape of the model curves remain constant throughout the whole data series\footnote{For more accurate modelling, apsidal motion should be taken into account.}.
For the initial assessment of $P$, we used a phase dispersion minimisation method \citep{Stellingwerf:1978}.
This initial period was necessary to fold the photometric and RV data for input to the modelling software. 
For the modelling we used PHysics Of Eclipsing BinariEs (\href{http://phoebe-project.org/}{PHOEBE}), (legacy version 0.31a), a binary modelling software based on the Wilson-Devinney code \citep{Wilson:1971}. 
The modelling was carried out using RV data from \citet{Kiminki:2008,Kiminki:2009,Kiminki:2012,Bohannan:1976,Rauw:1999} and photometric data from our 2011--2013 data set.
We used this particular photometry data set as it had relatively smaller errors and included all of our selected stars.
The spectral classes for the objects were taken from \citet{Kiminki:2014} and they were used as an input for the initial temperature \citep{Enc:2001} and mass values \citep{Martins:2005} for the models.
The resulting model parameters for masses ($M_1, M_2$), mass ratio ($q$), radii ($R_1, R_2$), temperatures ($T_1, T_2)$ and orbital parameters are presented in Table \ref{results_stellar}.

The phased curves from PHOEBE software were used as the input to the Bayesian inference tool called MultiNest \citep{Feroz:2008,Feroz:2009,Feroz:2013}.
The Bayesian parameter estimation method is a Monte Carlo algorithm that calculates the ephemerides of the star as 
\begin{equation}\label{hjd} \mathrm{HJD}=\mathrm{HJD_0}+PE+\frac{1}{2}\dot{P}PE^2, \end{equation}
where $E$ is the cycle number.

Using a random walk approach, we assessed the values of $P$ and $\dot{P}$ that best set the data to the model curve.
By comparing the folded light curve and folded RV curve that were calculated using the ephemerides to the model curves, we get the likelihood of the parameter fit.
The global maximum of the likelihood posterior is the best fit of the parameters in conjunction with the model curves.
We assumed the posterior to be Gaussian and thus the parameter fit errors are given as a standard deviation.
To minimise the errors, we fixed the value of $\mathrm{HJD_0}$ using the minimum of the modelled PHOEBE light curve.

For the period change analysis, we used one model light curve for all the photometric filters as the fitting of the ephemerides depends mainly on the time domain as opposed to the small differences in eclipse depths from different filters.

To test whether the PHOEBE model phased light curve causes any systematic errors in the period change determination, we constructed an empirical model light curves for our selected stars.
All the results for $P$ and $\dot{P}$ were obtained using both the PHOEBE model and the empirical model.
The empirical phased light curve which was sigma clipped and smoothed, was derived from all the photometric measurements where the first estimation of the period change was already accounted for.
As there were not enough RV data distributed over the full orbital phase, no empirical RV model curve could be accurately made.

The errors for each photometric data set were obtained separately as the standard deviation of the difference between the data points and the empirical light curve.
The errors for RV data were taken from the respective papers and the errors for the times of minimum in \citet{Sazonov:1961} were given as the timing error of the observations.
These errors determined the weight of each data set in the final period change analysis.

Conventionally, $O-C$ diagrams are constructed using only precisely measured photometric times of the eclipses where the times are obtained by continuous monitoring of the selected star during an eclipse.
Any RV data are usually completely ignored or only the times of conjunctions are used in the $O-C$ analysis.
In our case though, RV data can be fitted together with any photometric data.
This method is especially useful if one has sparsely covered time series of the eclipses and does not know the exact times of the minima.
This can happen either due to observational difficulties or when working with older data that have not been optimised in time domain for these kind of studies.
In general, our method can be used for any data coming from all-sky-covering campaigns.
They usually provide unevenly spaced observation epochs that make it harder to detect the times of minimum but on the other hand have a long time-series for a lot of objects.
Our simultaneous fitting of data with the appropriate weights provides a solution that is not dependent on the measurement errors of any single minimum and also improves the statistical significance of the period assessment as more data points are used.
Previous works for finding period change using all of the available data have been carried out for $\mathrm{RR_d}$ type stars \citep{Paparo:1998} and for selected eclipsing binaries \citep{Wilson:2014}.

\section{Results}

\begin{table*}
\caption{Orbital ephemerides from the Bayesian parameter estimation algorithm.}
\label{results_period}
\small
\centering
\begin{tabular}{l D{,}{~\pm~}{-1} D{,}{~\pm~}{-1} D{,}{~\pm~}{-1} D{,}{~\pm~}{-1} D{,}{~\pm~}{-1} D{,}{~\pm~}{-1} D{,}{~\pm~}{-1}}
\hline \hline
\rule{0pt}{8pt}Parameter & \multicolumn{1}{c}{A36} & \multicolumn{1}{c}{B17} & \multicolumn{1}{c}{MT059} & \multicolumn{1}{c}{MT696} & \multicolumn{1}{c}{MT720} & \multicolumn{1}{c}{Schulte~3} & \multicolumn{1}{c}{Schulte~5} \\
 \hline
\rule{0pt}{10pt}$\mathrm{HJD_0}$	  		   & \multicolumn{1}{c}{56282.520}      	& \multicolumn{1}{c}{56383.961} 	& \multicolumn{1}{c}{56535.672} 	        &\multicolumn{1}{c}{56181.719} 	       & \multicolumn{1}{c}{56585.602} 	& \multicolumn{1}{c}{56264.633}       & \multicolumn{1}{c}{56182.906} \\
$P_\mathrm{emp}~\mathrm{(d)}$   & \multicolumn{1}{c}{4.676040(21)}   	& \multicolumn{1}{c}{4.021767(38)} 	& \multicolumn{1}{c}{4.852281(39)} & \multicolumn{1}{c}{1.46918869(22)}    & \multicolumn{1}{c}{4.3619709(98)}    & \multicolumn{1}{c}{4.745964(15)}   & \multicolumn{1}{c}{6.597989(24)}\\
$\dot{P}_\mathrm{emp}~(\times~10^{-8})$ & 3.8,1.3 	& 1.1,1.6 	& -4.7,1.7   	& -0.37,0.08 & 8.1,2.1   & -1.5,1.0 & 1.16, 0.03\\
$P_\mathrm{model}~\mathrm{(d)}$	 & \multicolumn{1}{c}{4.676040(19)}   	& \multicolumn{1}{c}{4.021719(39)} 	& \multicolumn{1}{c}{4.852265(26)} 	& \multicolumn{1}{c}{1.46918857(24)}    &\multicolumn{1}{c}{4.3617773(96)}    & \multicolumn{1}{c}{4.745998(29)}   & \multicolumn{1}{c}{6.5980120(30)}\\
$\dot{P}_\mathrm{model}~(\times~10^{-8})$ & 3.9,1.3 	& -0.8,1.7 	& -3.6,1.2   	& -0.46, 0.08 & -11.7,3.2   & 0.0,0.9 & 1.30,0.03\\
\hline
\end{tabular}
\tablefoot{
Listed in this table are the epoch of periastron ($\mathrm{HJD_0}$), period ($P$) and the rate of period change ($\dot{P}$).
The parameters $P$ and $\dot{P}$ are calculated using empirical (emp) and PHOEBE (model) phased light curves.
$\mathrm{HJD_0}$ is fixed in both cases.
}
\end{table*}

We present the estimated orbital ephemerides in Table \ref{results_period} and binary parameters in Table \ref{results_stellar}. 
We give one-sigma likelihood-estimation uncertainties for $P$ and $\dot{P}$, formal errors from PHOEBE for $T_2, i, e$ and calculated formal errors\footnote{http://phoebe-project.org/1.0/docs/phoebe\_manual.pdf} for $M_1, M_2, R_1, R_2$ and $a$.

Our approach in this work does not take into account any cyclic patterns of period changes which may come from the light travel time effect through the presence of a third body in the system or from the stellar magnetic activity cycles \citep{Liao:2010}.
As data for most of our selected stars have a relatively short time base compared to the typical periods of the third orbiting body or magnetic cycles, we have opted to not include a cyclic pattern as a free parameter because this would artificially introduce a new degeneracy to the fit.
So any cyclic patterns in the period change are sources of possible errors in our results.

Also, we have not taken into account the apsidal motion in binaries with eccentric orbits (MT720 and MT059). 
The apsidal motion would visibly alter the shape of light and RV curves but we did not see this in our data sets (see Appendix \ref{sec:app}).

\begin{figure}[t]
	\includegraphics{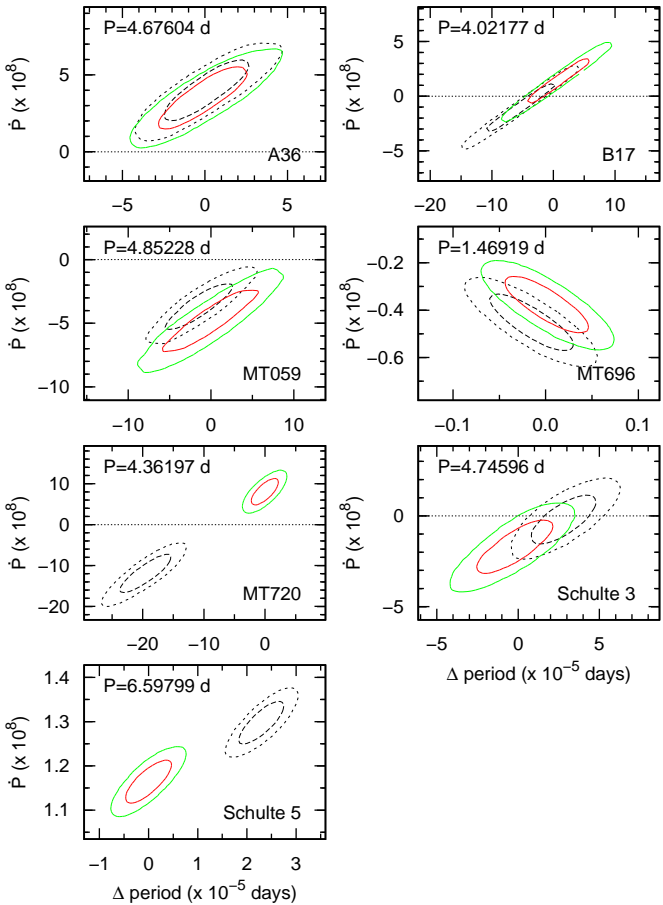}
	\caption{The parameter estimation results for period and period change. Periods are given as a difference from the $P_\mathrm{emp}$ value (see Table~\ref{results_period}). Coloured solid line represents the results for an empirical phase model and dashed line represents the results for calculated PHOEBE phase model. Inner contour is the 68\% and outer contour is the 95\% confidence level.}
	\label{p-pd}
\end{figure}

Figure~\ref{p-pd} shows the parameter $P$ and $\dot{P}$ estimation results for both empirical and PHOEBE model folded light curves.
The shape of the resulting parameter estimation likelihood cloud shows the amount of degeneracy of the two parameters as well as the uncertainties in the parameter estimations.
Due to the degeneracy of $P$ and $\dot{P}$, the given uncertainties for those parameters are determined by marginalised distributions.
If we had also fitted the $\mathrm{HJD_0}$ value, the results would have been degenerated between all three of those parameters.
Therefore, the value of $\mathrm{HJD_0}$ was fixed to reduce the systematic errors of the two other fitted parameters that we are interested in.
The distance between the locations of the parameter estimation likelihood clouds shows the systematic errors that arise from different model curve shapes and the sizes of those clouds show the statistical errors.
So one has to pay attention to the model folded light curves as they introduce some systematic errors compared to the true light curve.

Appendix \ref{sec:app} describes the model fits for each of the dataset separately.
The visual comparison between the data and the models can be used to assess the applicability of the final resulting parameters.

By comparing the results to each other from the empirical and PHOEBE folded light curve models for each star in Fig.~\ref{p-pd}, the results look consistent for most cases.
The results for Schulte~5 seem to disagree with each other but, due to small uncertainties in the parameter estimation, are actually in good agreement.
The resulting uncertainties for Schulte~5 are very small because its time base is the longest out of our selected objects.
The results for MT720 show a very large systematic error between the two models.
This is due to the relatively short time base as well as the difference between the shape of those two model folded light curves (see also Fig.~\ref{MT720}).

For four stars, we detect a non-zero period change values.
The strongest evidence for period change is found for Schulte~5, while MT696 also shows period change at three-sigma confidence level and A36 and MT059 at two-sigma confidence level.
MT059 and MT696 exhibit a negative period change, while A36 and Schulte~5 have a positive period change.
Our calculations for B17, MT720 and Schulte~3 do not indicate any change in their periods.

The stellar parameters (Table~\ref{results_stellar}) are obtained with the PHOEBE software where RV data are fitted alongside photometric data.
The stellar parameter values are not fine-tuned for each individual star as they are a by-product of acquiring model phase curves.
As the parameter modelling is degenerate between many of the parameters, adding high-resolution spectra for precise temperature and spectral-type measurements would constrain the models further.

\begin{figure}[t]
	\includegraphics{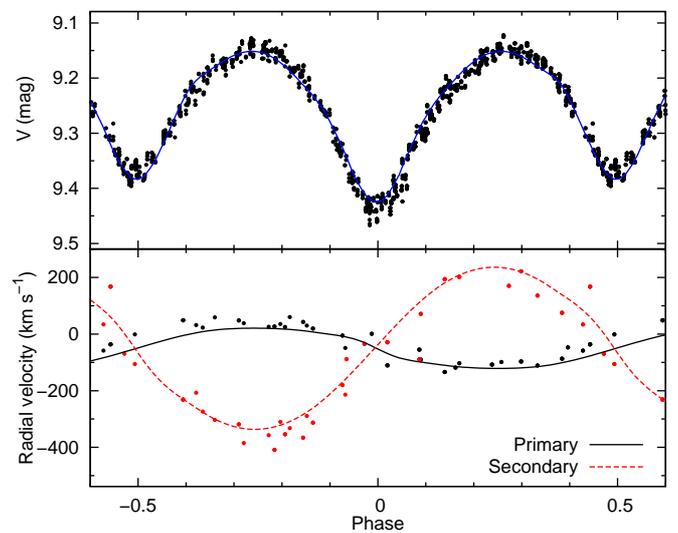}
	\centering
	\caption{The binary modelling for Schulte~5 with the PHOEBE software. Upper panel shows our photometric iTelescope data (data set 1) with the model curve (solid line) and lower panel shows combined radial-velocity data from \citet{Bohannan:1976} and \citet{Rauw:1999} with the model curves for both primary and secondary components.}
	\label{phoebe_schulte5}
\end{figure}

\begin{figure}[t]
	\includegraphics{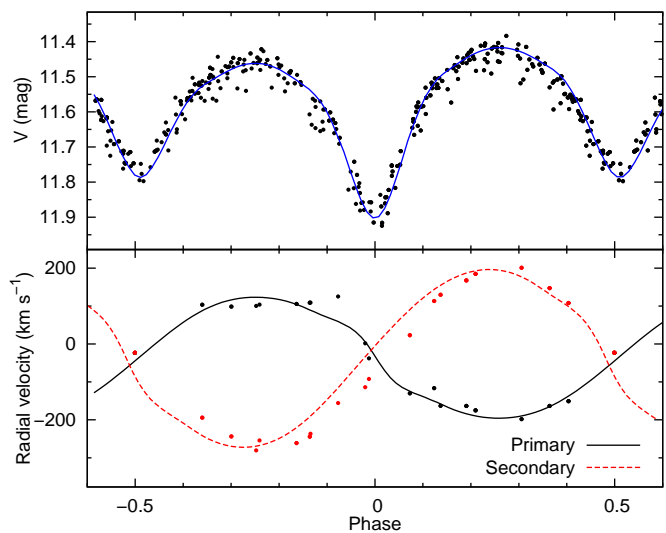}
	\centering
	\caption{The binary modelling for A36 with the PHOEBE software. Designations of lines, labels and photometric data set are the same as in Fig.~\ref{phoebe_schulte5}. Radial-velocity data are taken from  \citet{Kiminki:2009}.}
	\label{phoebe_A36}
\end{figure}

\begin{figure}[t]
	\includegraphics{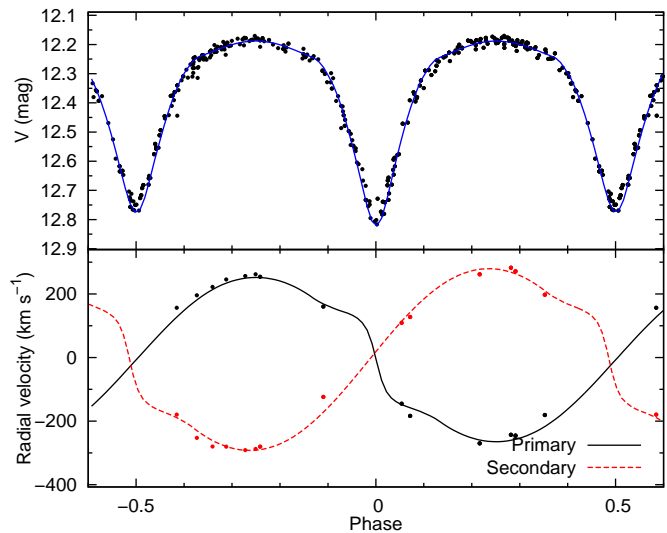}
	\centering
	\caption{The binary modelling for MT696 with the PHOEBE software. Designations of lines, labels and photometric data set are the same as in Fig.~\ref{phoebe_schulte5}. Radial-velocity data are taken from  \citet{Kiminki:2012}.}
	\label{phoebe_MT696}
\end{figure}

\begin{figure}[t]
	\includegraphics{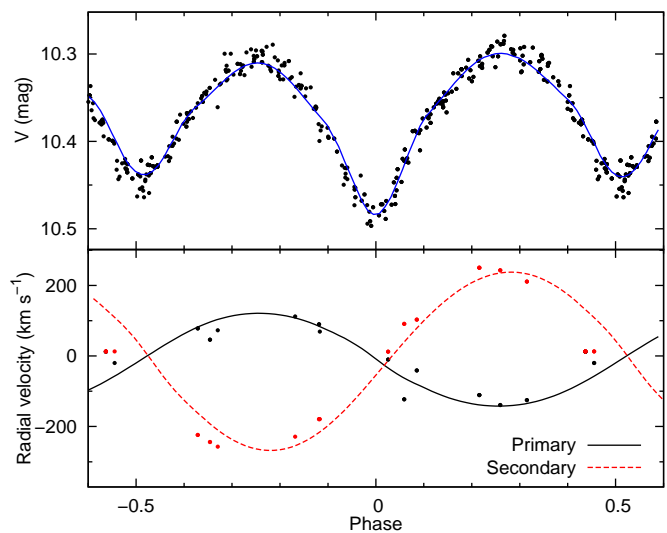}
	\centering
	\caption{The binary modelling for Schulte~3 with the PHOEBE software. Designations of lines, labels and photometric data set are the same as in Fig.~\ref{phoebe_schulte5}. Radial-velocity data are taken from  \citet{Kiminki:2008}.}
	\label{phoebe_schulte3}
\end{figure}

There are no parameters for B17 because the only available RV data from \citet{Comeron:2002} are published as a phased graph with no available ephemerides.
Without RV data, no orbital parameters or masses can be accurately modelled.
Also, for MT059, the RV data are only available for the primary component and thus the model stellar parameters are only one of the possible parameter fits due to the degeneracy of the mass ratio.
All the fits of our data to the acquired binary models are presented in Figs.~\ref{phoebe_schulte5}~--~\ref{phoebe_schulte3}.

\begin{table*}
\caption{Binary parameters from the PHOEBE modelling software.}
\label{results_stellar}
\footnotesize
\centering
\begin{tabular}{l D{,}{~\pm~}{5} c D{,}{~\pm~}{3} D{,}{~\pm~}{3} D{,}{~\pm~}{6} D{,}{~\pm~}{4} D{,}{~\pm~}{8}}
\hline \hline
\rule{0pt}{8pt}Parameter & \multicolumn{1}{c}{A36} & \multicolumn{1}{c}{B17} & \multicolumn{1}{c}{MT059} & \multicolumn{1}{c}{MT696} & \multicolumn{1}{c}{MT720} & \multicolumn{1}{c}{Schulte~3} & \multicolumn{1}{c}{Schulte~5} \\
 \hline
\rule{0pt}{8pt}$M_1$ $(M_{\odot})$ 	   & 22.41,2.96  &	--	& 21.23,6.24 	& 14.00,0.81     & 18.52,1.51 	& 32.73,7.04 	  & 38.18,5.21 \\
$M_2$ $(M_{\odot})$ 	   & 15.32,2.02 &	--	& 6.22,1.83 	& 12.67,0.73     & 13.19,1.07 	& 17.21,3.70 	  & 10.24,1.40\\
$q$ 			   & 0.68,0.04 	&	--	& 0.29,0.01	& 0.91,0.03      & 0.71,0.03  	& 0.53,0.05 	  & 0.27,0.03 \\
$R_1$ $(R_{\odot})$ 	   & 16.08,0.27	&	--	& 12.51,0.10 	& 5.92,0.09      & 9.93,0.59 	& 18.40,0.41 	  & 27.16,0.34\\
$R_2$ $(R_{\odot})$ 	   & 13.48,0.30	&	--	& 5.89,0.21 	& 5.63,0.09      & 8.47,0.38 	& 13.50,0.34	  & 15.15,0.20\\
$T_1$ $(\mathrm{K})$ (fixed) 	   & \multicolumn{1}{c}{27000}	&--&\multicolumn{1}{c}{38000}& \multicolumn{1}{c}{32000}      & \multicolumn{1}{c}{29800} & \multicolumn{1}{c}{41200}  & \multicolumn{1}{c}{36100}\\
$T_2$ $(\mathrm{K})$ 	   & 22924,38 	&	--	& 18966,98	& 31070,25 	 & 19202,80     & 29702,37 	  & 23681,112\\
$i$ $(\mathrm{^{\circ}})$   & 71.76,0.11 &	--	& 68.18,0.13 	& 81.87,0.06     & 71.09,0.28 	& 57.47,0.09 	  & 64.65,0.05 \\
$a$ $(R_{\odot})$ 	   & 39.43,1.45 &	--	& 36.34,3.46 	& 16.23,0.24     & 35.52,0.78 	& 43.72,2.63 	  & 53.90,2.08\\
$e$ 			   & 0.01,0.001 &	--	& 0.14,0.002 	& \multicolumn{1}{c}{0 (fixed)}	       & 0.34,0.005 	& 0.015,0.001 	  & \multicolumn{1}{c}{0 (fixed)}\\
$\omega$ $(\mathrm{^{\circ}})$ 	 & \multicolumn{1}{c}{350}    & 	--	& \multicolumn{1}{c}{189} & \multicolumn{1}{c}{0 (fixed)}     & \multicolumn{1}{c}{298}	& \multicolumn{1}{c}{0}   & \multicolumn{1}{c}{0 (fixed)}\\
Spectral class & \multicolumn{1}{c}{B0\RNum{1}b+B0\RNum{3}} & O7:+O9: & \multicolumn{1}{c}{O8\RNum{5}+B} & \multicolumn{1}{c}{O9.5\RNum{5}+B0\RNum{5}} & \multicolumn{1}{c}{B0-B1\RNum{5}+B1-B2\RNum{5}} & \multicolumn{1}{c}{O6\RNum{4}:+O9\RNum{3}} & \multicolumn{1}{c}{O7\RNum{1}anfp+Ofpe/WN9} \\
\hline
\end{tabular}
\tablefoot{Parameters in this table are the primary and the secondary component masses ($M_1$, $M_2$), the mass ratio $M_1/M_2$ ($q$), the primary and the secondary component radii ($R_1$, $R_2$), the primary and the secondary component temperatures ($T_1$, $T_2$), the inclination angle ($i$), the semi-major axis ($a$), the eccentricity of the system ($e$), the argument of periastron ($\omega$) and the binary spectral classes that are taken from \citet{Kiminki:2014}. The temperature for the primary component was fixed according to the spectral class \citep{Enc:2001}.
}
\end{table*}

\section{Discussion}

We use our estimated period change values to derive mass-loss rates for two distinctive cases.
If we assume a conservative mass exchange between the two components with constant system orbital angular momentum and system mass, the mass transfer rate can be calculated as case~\RNum{1} in \citet{Singh:1986}: 

\begin{equation}\label{case1} \dot{M_1}=\frac{1}{3} \frac{M_1 M_2}{M_1-M_2}\frac{\dot{P}}{P}. \end{equation}

We can also assume that the mass loss is solely through stellar wind \citep[case~\RNum{4} in][]{Singh:1986}:

\begin{equation}\label{case4} \dot{M_1}=-\frac{1}{2} (M_1+M_2)\frac{\dot{P}}{P}. \end{equation}

\begin{table}
\caption{The mass transfer and mass loss through stellar wind rates for our selected stars. No mass loss through stellar wind is given for the negative $\dot{P}$ value as this would indicate a positive mass loss from the primary. The errors are calculated using the $P$, $\dot{P}$, $M_1$ and $M_2$ uncertainties. The errors for MT696 are large due to the similar companion masses.}
\label{mass-loss}
\centering
\begin{tabular}{ll D{,}{~\pm~}{6} D{,}{~\pm~}{6}}
\hline \hline
\rule{0pt}{10pt}Star	&used $\dot{P}$& \multicolumn{1}{c}{$\dot{M_1}$ mass transfer}	& \multicolumn{1}{c}{$\dot{M_1}$ stellar wind} \\
	&value&\multicolumn{1}{c}{($10^{-6}~M_{\odot}\mathrm{yr^{-1}}$)}& \multicolumn{1}{c}{($10^{-6}~M_{\odot}\mathrm{yr^{-1}}$)} \\
 \hline
\rule{0pt}{10pt}A36      &$\dot{P}_\mathrm{emp}$& 48, 19 &  -56, 19  \\
	 & $\dot{P}_\mathrm{model}$& 50 , 21  &  -58, 19  \\
MT059    &$\dot{P}_\mathrm{emp}$& -10.5, 4  & \multicolumn{1}{c}{--} \\
	 & $\dot{P}_\mathrm{model}$& -8.0 , 3   & \multicolumn{1}{c}{--}  \\
MT696    &$\dot{P}_\mathrm{emp}$& -41 , 160  &  \multicolumn{1}{c}{--} \\
	 & $\dot{P}_\mathrm{model}$& -50 , 254  &   \multicolumn{1}{c}{--}\\
MT720  	 &$\dot{P}_\mathrm{emp}$& 104 , 30 &  -10.7, 4.2 \\
	 & $\dot{P}_\mathrm{model}$& -149 ,45   &  \multicolumn{1}{c}{--} \\
Schulte~3&$\dot{P}_\mathrm{emp}$& -13.9 , 10.2   &  \multicolumn{1}{c}{--} \\
	& $\dot{P}_\mathrm{model}$ & 0.001 , 8.8  &   \multicolumn{1}{c}{--}\\
Schulte~5&$\dot{P}_\mathrm{emp}$& 3.0 , 0.2  &  -15.6,  0.7 \\
	 & $\dot{P}_\mathrm{model}$& 3.4 , 0.2 &  -17.4, 0.8 \\

\hline
\end{tabular}
\tablefoot{The two period change values ($\dot{P}_\mathrm{emp}$ and $\dot{P}_\mathrm{model}$) for each star are given in Table \ref{results_period}. B17 is omitted from the Table as there are no mass estimates.
}
\end{table}

Table~\ref{mass-loss} summarises for all of our selected stars the primary component mass-loss rates calculated for both conservative mass exchange and mass loss through stellar wind using the two acquired period change values of $\dot{P}_\mathrm{emp}$ and $\dot{P}_\mathrm{model}$.
Equations (\ref{case1}) and (\ref{case4}) are for the idealised occurrences where no other effect interferes with the period-change value.
The primary component mass-loss values in Table~\ref{mass-loss} represent isolated cases through either conservative mass exchange or stellar wind.
They do not indicate the real values of the mass loss as in reality the mass loss is a combination of different factors.
To constrain the mass loss values further, some additional method to find the strength of stellar wind (radio observations or analysis of the stellar-wind-induced spectral line profiles) has to be included.

In the next sections we compare our results to the previous findings as well as present some of the problems that we came across on a star by star case.

\subsection{Schulte~5}\label{sec:Schulte5}

Schulte~5 (V729 Cyg, BD +40$^{\circ}$ 4220) is an eclipsing binary of two O-type supergiants \citep{Kiminki:2014}.
There is also evidence for a third non-thermal radio source around the binary with a period of 6.7 years \citep{Kennedy:2010,Cazorla:2014}.

\citet{Linder:2009} assumed the orbital solution from \citet{Rauw:1999} to get the absolute masses of the components ($M_1=31.9~\pm~3.2~M_{\odot}$ and $M_2=9.6~\pm~1.1~M_{\odot}$).
Our modelling reveals similar masses for the components, $38.2~\pm~5.2~M_{\odot}$ and $10.2~\pm~1.4~M_{\odot}$ accordingly.

\citet{Martins:2005} have calculated a grid of models for O type stars using CMFGEN \citep{Hillier:1998} atmosphere code and they include theoretical spectroscopic masses for them.
For the primary component of spectral type O7\RNum{1}, the spectroscopic mass would be $M_{\mathrm{spec}}=40.91~M_{\odot}$.
They note that one should take care in dealing with their spectroscopic modelled masses as the uncertainty may be as high as 35 to 50\%.
Nevertheless, our resulting mass for the primary is more close to the theoretical value than to the result from \citet{Linder:2009}.

We chose Schulte~5 as a testbed for our period change study as the eclipse timing method for period change as well as the mass-loss rates have been studied on multiple occasions.
Previous assessments for the mass-loss rates are by \citet{Persi:1990} derived from infrared observations $\dot{M}=2.5~\times~10^{-5}~M_{\odot}\mathrm{yr^{-1}}$, by \citet{Conti:1999} combined with multi frequency radio observations $\dot{M}=(3.7~\pm~1.3)~\times~10^{-5}~M_{\odot}\mathrm{yr^{-1}}$ and by \citet{Linder:2009} using the period change rate of observed times of minimum $\dot{M}=(2.1~\pm~0.6)~\times~10^{-5}~M_{\odot}yr^{-1}$.

Following the calculation from Equation~(\ref{case1}), \citet{Linder:2009} found the mass-loss rate $\dot{M_1}=(4.7~\pm~1.5)~\times~10^{-6}~M_{\odot}\mathrm{yr^{-1}}$ which they argued against as this value was lower than expected from the previous assessments derived from the infrared and radio data. 
Using the masses obtained from our PHOEBE model and the $P$ and $\dot{P}$ from our Bayesian parameter estimation method, we also get similar mass-loss rates of $\dot{M_1}=(3.0~\pm~0.2)~\times~10^{-6}~M_{\odot}\mathrm{yr^{-1}}$ and $\dot{M_1}=(3.4~\pm~0.2)~\times~10^{-6}~M_{\odot}\mathrm{yr^{-1}}$ depending on the applied period change value.
This difference between \citet{Linder:2009}  $\dot{M_1}$ value and our values is mostly due to the difference in the period change values.
The value from \citet{Linder:2009}, $\dot{P}=(1.9~\pm~0.5)~\times~10^{-8}$, is higher than our obtained value of $\dot{P}_\mathrm{emp}=(1.16~\pm~0.03)~\times~10^{-8}$ and thus their mass-loss rate is also a little bit higher.
Our period change value should be more accurate as we have over 40 years longer time base.

\citet{Yakut:2014} found a period change of $P/\dot{P}~\sim~9.4~\times~10^5$ from eclipse timings by acquiring new observations on 65 observing nights from 2010 to 2011 in Bessel $U\!BV\!RI$ bands.
With their given period of $P=6.597981~\mathrm{d}$, we can calculate $\dot{P}=7.01913~\times~10^{-6}$, which is much higher from the value of $\dot{P}=(1.16~\pm~0.03)~\times~10^{-8}$ that we obtained.
It may be possible that their given $P/\dot{P}$ value is in the units of years instead of days as one might assume from their period value.
In this case the value $\dot{P}=1.92305~\times~10^{-8}$ agrees with the value of $\dot{P}$ from \citet{Linder:2009}.
They also model the masses of the primary ($31.6~\pm~2.9~M_{\odot}$) and secondary ($8.8~\pm~0.3~M_{\odot}$) component.
If we use their data for calculating mass loss through mass transfer, we get $\dot{M_1}=4.325~\times10^{-6}~M_{\odot}\mathrm{yr^{-1}}$ which is very similar to the mass-loss rate that \citet{Linder:2009} obtained but differs from ours due to the value of period change and obtained component masses.

We find that the mass-loss rate through stellar wind (Eq.~\eqref{case4}) for Schulte~5 is $\dot{M_1}=(-1.56~\pm~0.07)~\times~10^{-5}~M_{\odot}\mathrm{yr^{-1}}$ or $\dot{M_1}=(-1.74~\pm~0.08)~\times~10^{-5}~M_{\odot}\mathrm{yr^{-1}}$ depending on the $\dot{P}$ value, which are comparable to the result of \citet{Linder:2009} ($\dot{M_1}=(-2.1~\pm~0.6)~\times~10^{-5}~M_{\odot}\mathrm{yr^{-1}}$).
The previous mass-loss rate assessments of the stellar wind from radio and infrared observations have confirmed the value to be close to our finding.
This indicates the period change of the system to be caused by the mass loss from stellar wind of the primary component instead of the mass transfer from the secondary component.

\subsection{A36}\label{sec:A36}

A36 was first named in a paper by \citet{Comeron:2002} where they listed it as a possible binary.
The eclipsing nature of A36 was found from NSVS data \citep{Otero:2008,Hoffman:2009} and it is classified as an Algol/$\beta$ Lyrae candidate.
\citet{Kiminki:2009} determine the spectral type as a B0\RNum{1}b+B0\RNum{3} system on the basis that the light curve and spectra show similar component temperatures and luminosities with differences in specific lines.

Our calculated mass ratio for the binary system is 0.66 and it is within error limits of \citet{Kiminki:2009} estimation of $0.70~\pm~0.06$.
They give a period of $P=4.6752~\pm~0.0007~\mathrm{d}$ that is slightly different than our result of $P=4.67604~\pm~0.00002~\mathrm{d}$.
This is mainly due to the lack of observations in the RVS for A36 to precisely determine the period.
We find the period change from the combination of two models to be $\dot{P}=(3.85~\pm~1.3)~\times~10^{-8}$, thus the mass loss by direct mass transfer would be  $\dot{M_1}=(4.9~\pm~2.0)~\times~10^{-5}~M_{\odot}\mathrm{yr^{-1}}$ and by stellar wind $\dot{M_1}=(-5.7~\pm~1.9)~\times~10^{-5}~M_{\odot}\mathrm{yr^{-1}}$.

The value of $\dot{M_1}$ by stellar wind is too high for a B0\RNum{1}b star as the maximum empirical estimates for mass-loss rates are typically lower than $10^{-5}~M_{\odot}\mathrm{yr^{-1}}$ \citep{Vink:2000,Puls:2006}.
Therefore, the change in period must be caused by the mass transfer from the secondary to the primary.
This also indicates that the secondary component has to be more evolved than the primary.
Although the spectral classification by \citet{Kiminki:2009} contradicts this conclusion, it may still be viable if the luminosity class of the primary component could be lowered.
Our PHOEBE modelling results for the stellar radii and masses also support the argument that the primary component is more close to the luminosity class \RNum{3}.

\subsection{B17}\label{sec:B17}

B17 was first discussed in a paper by \citet{Comeron:2002} where they discovered that it displays emission in the Brackett $\gamma$ line (the B notation in front of the star number) which is an indicator of high mass-loss rate, typical to the evolved stellar stages.
It is a massive evolved binary system with a spectral classification estimation of O7\RNum{1}a+O9\RNum{1} supergiants.
\citet{Hoffman:2009} classified it as an Algol/$\beta$ Lyrae type using NSVS data.

For B17 we could not use any spectroscopic data.
\citet{Stroud:2010} have done an extensive photometric and spectroscopic campaign on B17, but unfortunately their data are not publicly available and they are only in a phased form on a graph.
Their determined period of $4.02174~\pm~0.00003$~d coincides with ours and as the time base between the NSVS photometry data and their observed spectra was only four years, they do not discuss any possibility of a period change.
We did not find any indication of a period change within the binary system with values of  $\dot{P}_\mathrm{emp}=(1.1\pm 1.6)~\times~10^{-8}$ and $\dot{P}_\mathrm{model}=(-0.8\pm 1.7)~\times~10^{-8}$ and therefore we find no mass loss in the system.
Also due to no RV data available, no stellar parameters were modelled.

\subsection{MT059}\label{sec:MT059}

\begin{figure}[t]
	\includegraphics{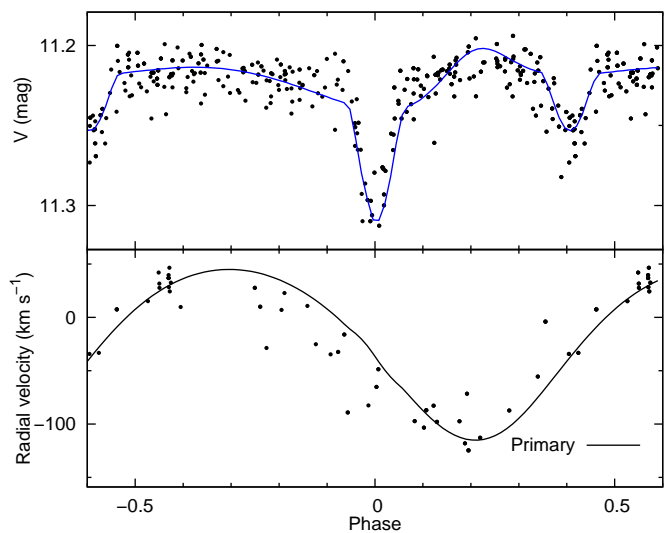}
	\centering
	\caption{The binary modelling for MT059 with the PHOEBE software.Designations of lines, labels and photometric data set are the same as in Fig.~\ref{phoebe_schulte5}. Radial-velocity data are taken from  \citet{Kiminki:2008}.}
	\label{phoebe_MT059}
\end{figure}

MT059 (Schulte 1) is O8V type star with a B, possibly later than B1V, type companion.
\citet{Kiminki:2008} have only measured radial velocities for the primary component and have not detected any spectral features of the companion thus indicating that the luminosity ratio has to be less than $L_2/L_1~\lesssim~0.2$.
For their primary component mass assessment, they use the theoretical mass for a spectral type O8\RNum{5} from \citet{Martins:2005} with a value of $M_{\mathrm{spec}}=21.95~M_{\odot}$.
They come to the conclusion that due to the absence of spectral signatures of the secondary component, the interpolation from its spectral type yields the mass of the secondary companion to be $5.1~M_{\odot}~\lesssim~M_2~\lesssim~13.8~M_{\odot}$ and the mass ratio $0.24~\pm~0.02~\lesssim~q~\lesssim~0.64~\pm~0.03$.
Our resulting mass ratio falls on a lower side of their prediction with a value of 0.29 and the masses we obtained are in agreement with the previously mentioned predictions with values of $M_1=21.23~\pm~6.24~M_{\odot}$ and $M_2=6.22~\pm~1.83~M_{\odot}$.
As there are no RV data for the secondary companion, we cannot assess the mass ratio independently and thus the mass ratio is degenerate with the other stellar parameters.
Also the semi-major axis of the system yields large formal errors which lead to large errors in the mass assessments.
The modelled light curve (upper panel in Fig.~\ref{phoebe_MT059}) indicates the presence of a reflection effect (phase $\sim0.2$) that is only loosely supported by the data.
A more thorough investigation is needed to determine whether this effect is from the observational uncertainties or from the underlying physics of the system.

MT059 is indicated by \citet{Kiminki:2008} as a SB1--type binary with no mentioning of its eclipsing nature.
Our observations, as well as NSVS photometric data, show clear eclipses from the photometric folded light curve.
The binary has an eccentricity of 0.14 which is also noticeable from the uneven timings in phase space between the primary and secondary eclipse.
  
We find that out of all of our selected binary systems, MT059 has the highest period change value ($\dot{P}_\mathrm{emp}=(-4.7~\pm~1.7)~\times~10^{-8}$ or $\dot{P}_\mathrm{model}=(-3.6~\pm~1.7)~\times 10^{-8}$).
This is due to the RV observations that are also visibly shifted in our PHOEBE modelling (Fig.~\ref{phoebe_MT059}).
The mass loss through conservative mass exchange with the results from our empirical model is $\dot{M_1}=(-10.5~\pm~4)~\times~10^{-6}~M_{\odot}\mathrm{yr^{-1}}$.
As the period change is negative, the mass-loss from stellar wind cannot be calculated using Eq.~\eqref{case4} because it would result in a positive $\dot{M_1}$ value and therefore the period change must be from mass transfer from the primary to the secondary component.

\subsection{MT696}\label{sec:MT696}

MT696 is a binary system of an O9.5V spectral type primary and an early B type companion.
\citet{Rios:2004} were the first to report the eclipsing binary status of MT696, indicating that it might be a W UMa type binary with a 1.46-day period.
\citet{Kiminki:2008} argue that the early spectral types of the components, the absence of emission features, and a period over 1 day suggest MT696 to be $\beta$ Lyr type instead.
They also improve the period estimation to $1.4692~\pm~0.0005$ days and estimate the inclination to be near $80^{\circ}$ and the mass ratio $q$ to be $0.85~\pm~0.03$, which are close to our estimations.
Their minimum mass estimations are $15.1~\pm~0.7~M_{\odot}$ for the primary and $12.8~\pm~0.5~M_{\odot}$ for the secondary component, while our modelling gives the total masses of $14.00~\pm~0.81~M_{\odot}$ and $12.67~\pm~0.73 M_{\odot}$, respectively.
From \citet{Martins:2005} the mass for spectral type O9.5\RNum{5} is $M_{\mathrm{spec}}=16.46~M_{\odot}$.
Our mass estimations are within errors of the minimum masses from \citet{Kiminki:2008} but lower than the theoretical primary mass, so our resulting masses might be underestimated.
Another possibility is that the spectral types and thus the masses of the binary components might not be accurately determined by \citet{Kiminki:2014}.

\citet{Souza:2014} find the period to be $1.46919~\pm~0.00006~\mathrm{d}$ and they did not find any period change in their two year long data set.
We, on the other hand, find the period change at three-sigma confidence level with the value of $\dot{P}_\mathrm{emp}=(-0.37~\pm~0.08)~\times~10^{-8}$ and $\dot{P}_\mathrm{model}=(-0.46~\pm~0.08)~\times~10^{-8}$.
This is the smallest period change value amongst our studied binaries.
Due to the small value for period change and the similar masses for the binary components, the errors for conservative mass-loss rates are very high, as the total error depends on the difference between the companion masses.
The acquired mass-loss rate for direct mass transfer is then $\dot{M_1}=(-0.4~\pm~1.6)~\times~10^{-4}~M_{\odot}\mathrm{yr^{-1}}$ and it is dominated by the measurement errors of the stellar masses.

\subsection{MT720}\label{sec:MT720}

\begin{figure}[t]
	\includegraphics{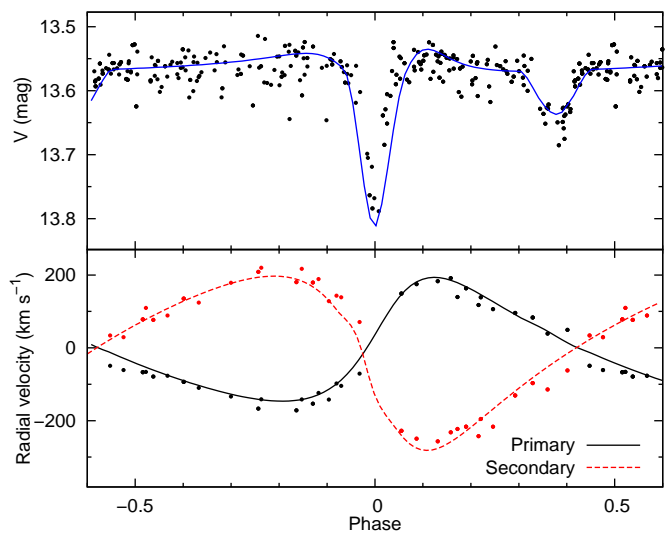}
	\centering
	\caption{The binary modelling for MT720 with the PHOEBE software. Designations of lines, labels and photometric data set are the same as in Fig.~\ref{phoebe_schulte5}. Radial-velocity data are taken from  \citet{Kiminki:2008}.}
	\label{phoebe_MT720}
\end{figure}

\citet{Kiminki:2012} have observed MT720 and measured its radial velocities on 32 occasions, noting that no known photometric variability have been studied.
They indicate the period of the system to be $4.3622~\pm~0.0003$ days (compared to our $P_{\mathrm{emp}}=4.36197~\pm~0.00001$ and $P_{\mathrm{model}}=4.36177~\pm~0.00001$ days) and an eccentricity of $0.35~\pm~0.02$, which is evident from the RV curve.
They also find minimum masses from the RV curve and suggest the spectral type of the binary system to be B0--B1\RNum{5} + B1--B2\RNum{5}.
Their calculated minimum masses of $M_1~\geq~15.5~M_{\odot}$ and $M_2~\geq~11.1~M_{\odot}$ are in accordance with our modelling results for the component total masses of $M_1=18.52~\pm~1.51~M_{\odot}$ and $M_2=13.19~\pm~1.07~M_{\odot}$.
 
\citet{Salas:2014} have done a 1.5-year-long photometric survey in Cygnus OB2 and have included MT720 as a confirmed eclipsing binary with a period of $4.3619~\pm~0.0001$ days.
Our photometric data also show clear evidence of eclipsing nature of the binary.
At first we note that in phase space the primary eclipse coincides with the RV curve when the more massive star is eclipsing the secondary component, indicating that it must be less luminous. 
After modelling the light curve, it turns out that because of the high eccentricity of the system, the primary eclipse depth lessens into what we see as a secondary eclipse.
This effect is due to the combination of the high eccentricity and a low inclination angle resulting in a partial eclipsing of the primary component by the secondary star.
The model in Fig.~\ref{phoebe_MT720} is acquired by introducing a phase shift of 0.45 so the secondary eclipse would coincide with a phase zero point.

Modelled stellar parameters for MT720 are more prone to errors compared to the other binary systems due to the bad fit between the data and the model.
As can be seen in Fig.~\ref{phoebe_MT720}, the secondary eclipse in the observations is narrower compared to the model curve.
Our period change modelling for this system have a strong systematic effect due to the differences between the two phased light curve models. 
The above mentioned modelling discord is due to the fact that MT720 have the largest photometric uncertainties as it is the faintest object in our selection.
Because of those systematic errors between the models, we cannot find any reliable evidence for the period change in MT720.

\subsection{Schulte~3}\label{sec:Schulte3}

Schulte~3 is a massive binary system consisting of O6\RNum{4}: and O9\RNum{3} stellar type components.
\citet{Kiminki:2008} found a mass ratio of $q=0.44~\pm~0.08$ which agrees within errors with our result of $0.53~\pm~0.05$ but differs from the theoretical mass ratio of $q=0.64$ for O6\RNum{4}: + O9\RNum{3} binary system \citep{Martins:2005}. 
We find our modelled masses ($M_1=32.73~\pm~7.04~M_{\odot}$ and $M_2=17.21~\pm~3.70~M_{\odot}$) to agree with their minimum mass predictions of $M_1~\geq~17.2~M_{\odot}$ and $M_2~\geq~7.6~M_{\odot}$.
Our acquired primary component mass is also close to the theoretical spectroscopic mass of O6\RNum{5} ($M_{\mathrm{spec}}=31.73~M_{\odot}$), but the theoretical mass value for the secondary component ($M_{\mathrm{spec}}=23.07~M_{\odot}$) is higher than our result.

\citet{Kiminki:2008} argued that the difference between the mass ratios is due to the mass transfer between the components.
The period we find coincides with their result ($P=4.7459~\pm~0.0003$ days) within the errors, but we do not detect any notable period change in the system.
Our period change value for empirical model is $\dot{P}_\mathrm{emp}=(-1.5~\pm~1.0)~\times~10^{-8}$, which does not imply a non-zero value at two-sigma confidence level, and for PHOEBE model the value is  $\dot{P}_\mathrm{model}=(0.0~\pm~0.9)~\times 10^{-8}$.
These values show clear evidence for no period change within the system and indicate no mass transfer between the components.

\section{Conclusions}\label{sec:Conclusions}

We studied the period change of seven massive binaries in the Cygnus OB2 association.
For this we used archival photometry and RV data together with our own photometric observations from 2011--2014.
Our observations were essential in the period-change analysis for A36, B17, MT696 and MT720 as they increased the time base up to 13 years.
We note for the first time the eclipsing nature of MT059.

Four stars show evidence of a non-zero period change at two-sigma confidence level.
For three stars we report the period change for the first time.
Our calculations confirm the previous assessments for $\dot{P}$ of a star Schulte~5.
We also show that the change in period in Schulte~5 originates from the stellar wind from the primary component.
Two of the systems, B17 and Schulte~3, show period change values close to zero indicating that there is no mass transfer between the binary components or that the period change values are too small for us to estimate them reliably with the available data.
Our method had systematic errors in fitting the period and period change parameters for MT720 and thus we did not derive any reliable values for that system.
Two of the systems, MT059 and MT696, have negative $\dot{P}$ values that show mass transfer from the primary to the secondary component.
We argue that the change in period for A36 might be from mass transfer from the secondary component to the primary, indicating a more evolved secondary stellar component.

We show that applying the Bayesian parameter estimation method, period change can be calculated using all the available data simultaneously for any binary star.
Utilising all data points in contrast to using only the times of minimum, improves the statistical significance of the calculations and opens up new data sets to include in the analysis.
This is especially important with the trend to digitise archives of photographic plates as it adds new historical observations that were not optimised for the eclipses of binaries, which means that the times of minimum may be difficult to obtain.

\begin{acknowledgements}
This work was supported by institutional research funding IUT40-1 and IUT40-2 of the Estonian Ministry of Education and Research, by the Estonian Research Council grants PUTJD5 and 8906. 
Authors thank Taavi Tuvi for his involvement in finding the literature data and doing the preliminary analysis for finding the period change values.
All the figures were made using the gnuplot plotting utility. In this work the SExtractor source extraction software was used \citep{Bertin:1996}.
\end{acknowledgements}

\bibliographystyle{aa} 
\bibliography{Laur2015} 
%\listofobjects
% \clearpage

%\Online
\begin{appendix}

\section{Individual data sets and model curves}\label{sec:app}

\begin{figure}[!ht]
	\includegraphics{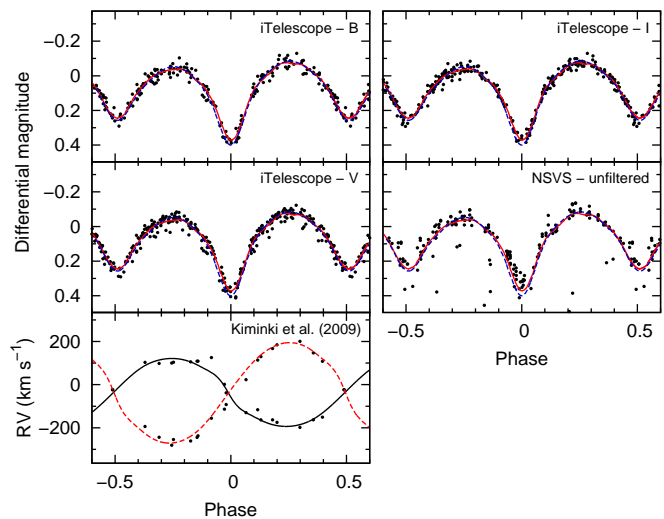}
	\centering
	\caption{Phase diagrams for A36 using the $P_\mathrm{emp}$ and $\dot{P}_\mathrm{emp}$ from Table \ref{results_period}. Blue dashed lines in the magnitude plots denote a model light curve from PHOEBE software and the red solid lines denote an empirical light curve. The solid line in the RV plot represents the model for the primary component and the red dashed line represents the model for the secondary component.}
	\label{A36}
\end{figure}

\begin{figure}[!ht]
	\includegraphics{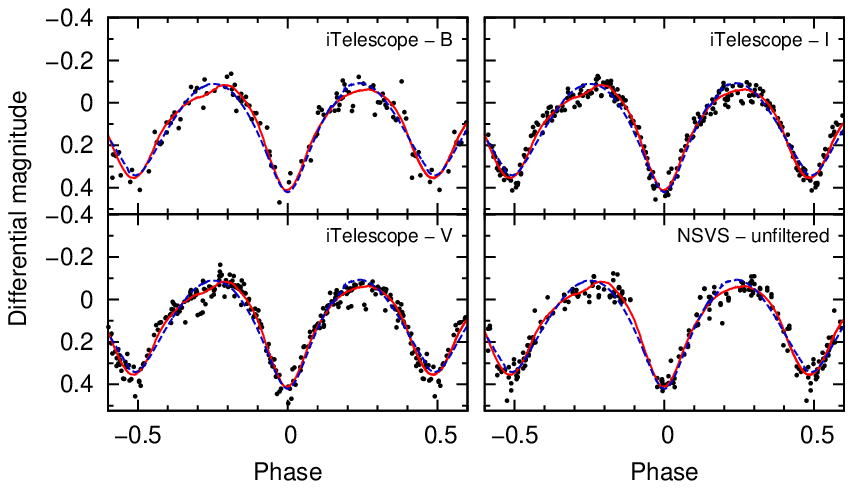}
	\centering
	\caption{The phase diagrams for B17. Designations are the same as in Fig. \ref{A36}.}
	\label{B17}
\end{figure}

\begin{figure}[!ht]
	\includegraphics{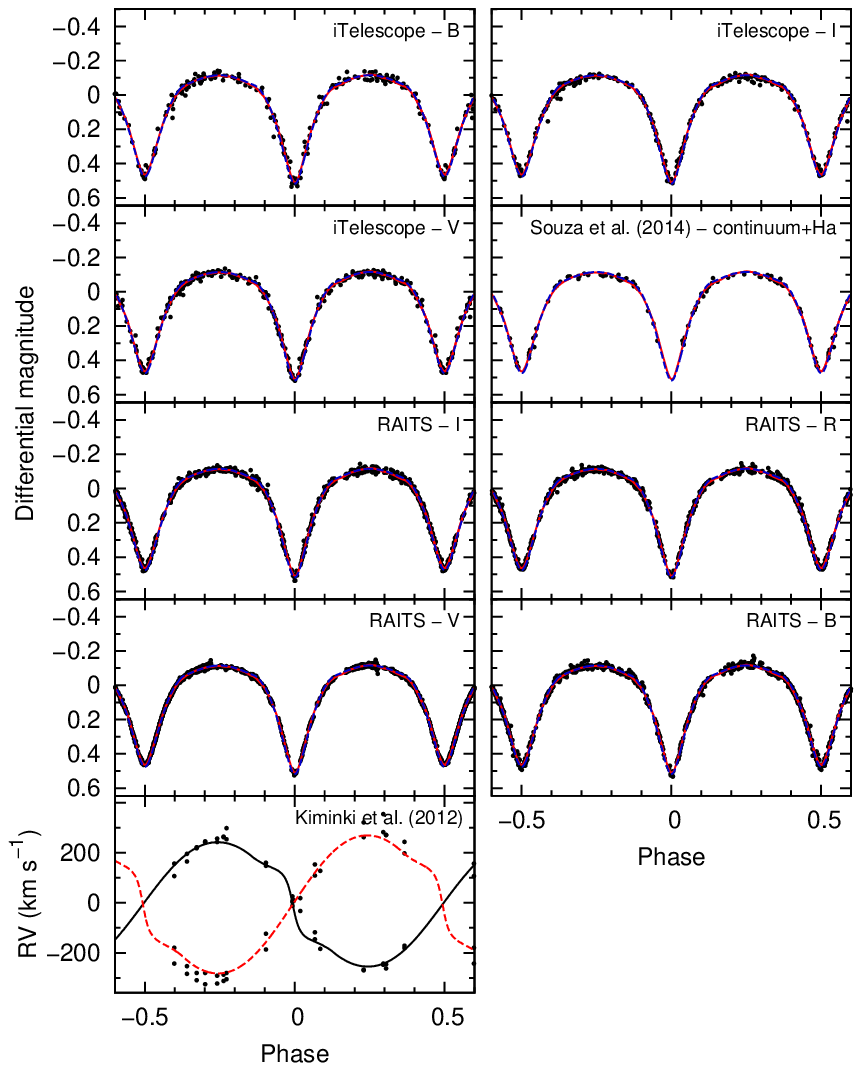}
	\centering
	\caption{The phase diagrams for MT696. Designations are the same as in Fig. \ref{A36}.}
	\label{MT696}
\end{figure}

\begin{figure}[!ht]
	\includegraphics{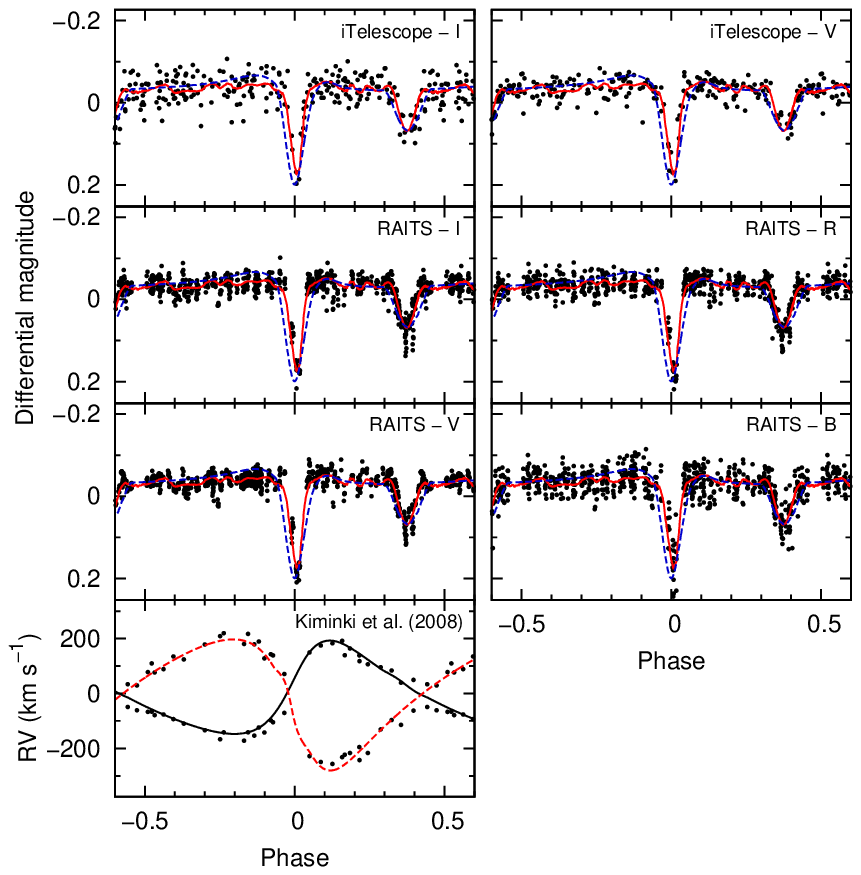}
	\centering
	\caption{The phase diagrams for MT720. Designations are the same as in Fig. \ref{A36}.}
	\label{MT720}
\end{figure}

\begin{figure}[!ht]
	\includegraphics{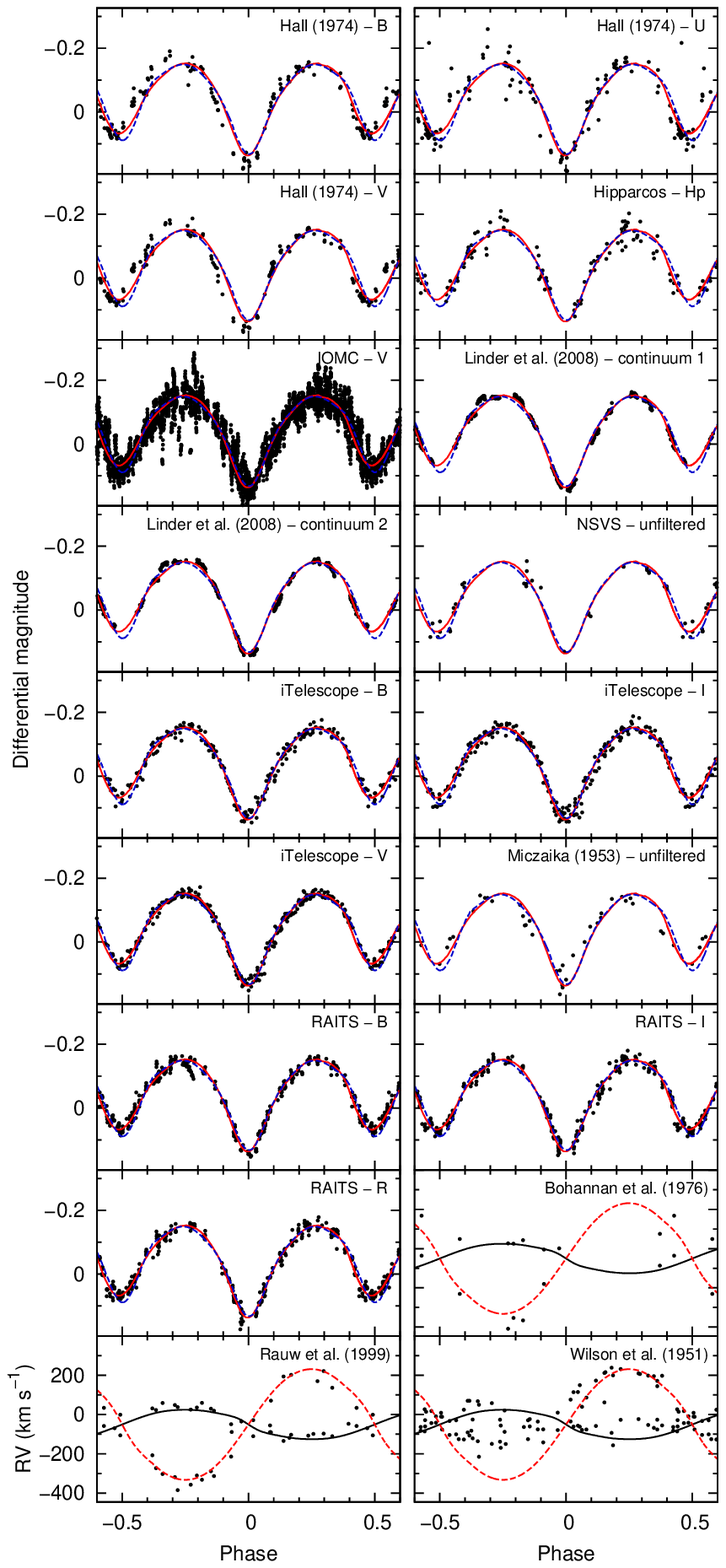}
	\centering
	\caption{The phase diagrams for Schulte~5. Designations are the same as in Fig. \ref{A36}.}
	\label{schulte5}
\end{figure}

\begin{figure}[!h]
	\includegraphics{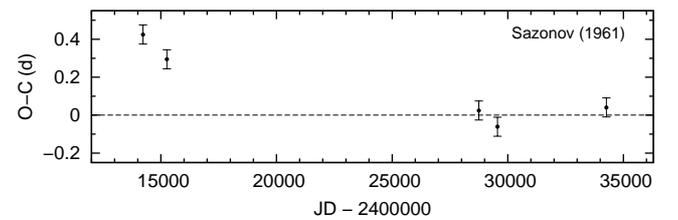}
	\centering
	\caption{The $O-C$ diagram for Schulte~5. Points denote the time difference between the observed times of minimum and the calculated times of minimum using the $P_\mathrm{emp}$ and $\dot{P}_\mathrm{emp}$ from Table \ref{results_period}. Dashed line shows the $O-C$ value of zero.}
	\label{min_schulte5}
\end{figure}

\begin{figure}[!ht]
	\includegraphics{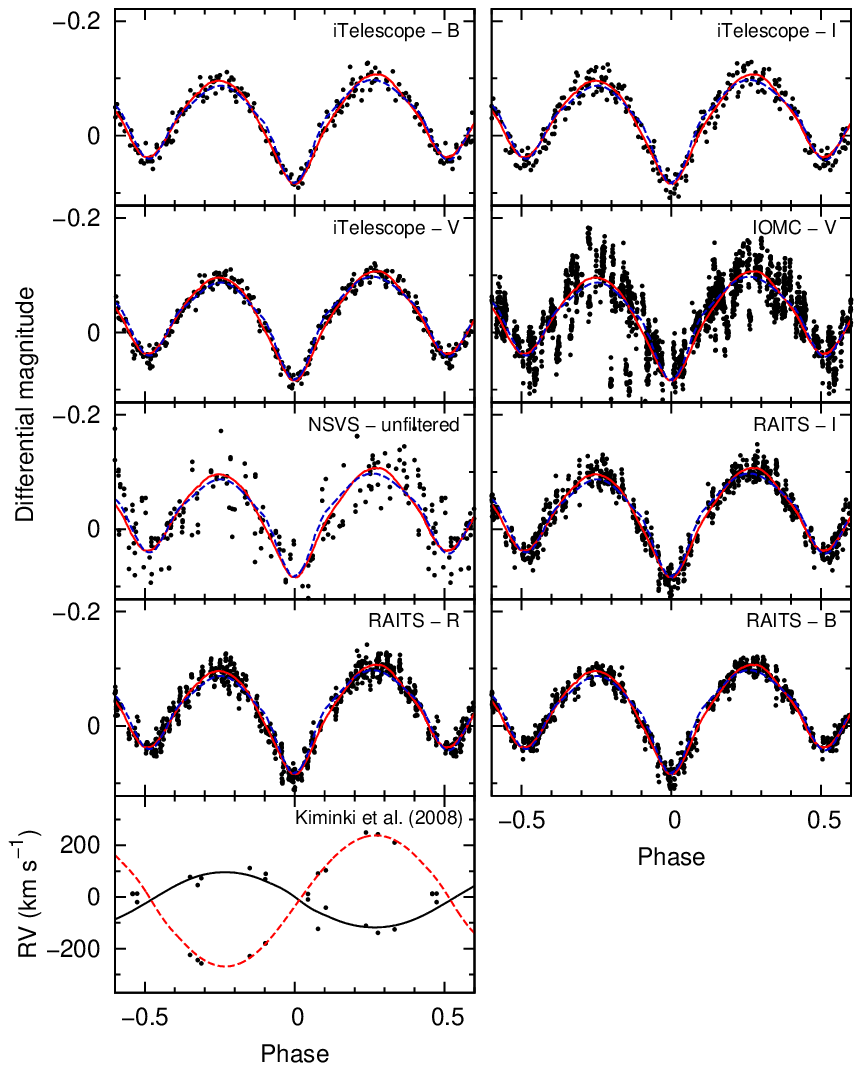}
	\centering
	\caption{The phase diagrams for Schulte~3. Designations are the same as in Fig. \ref{A36}.}
	\label{schulte3}
\end{figure}

\begin{figure}[!ht]
	\includegraphics{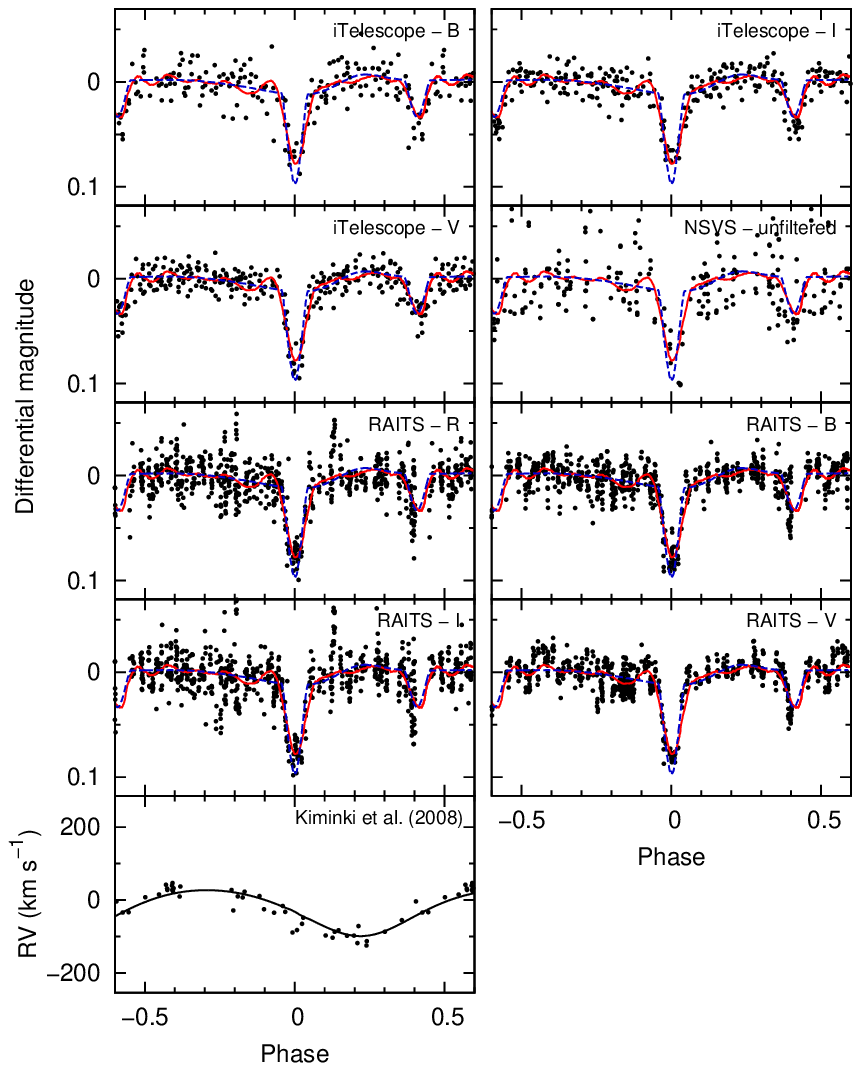}
	\centering
	\caption{The phase diagrams for MT059. Designations are the same as in Fig. \ref{A36}.}
	\label{MT059}
\end{figure}

In this section we present the resulting phase diagrams for photometric and RV data sets for our studied stars.
The diagrams are constructed by folding the data as Equation~(\ref{hjd}) with the resulting orbital ephemerides (Table~\ref{results_period}).
Figures~\ref{A36}~--~\ref{MT059} show individual data sets together with the phased model curves.

\end{appendix}
\end{document}